\documentclass[usenatbib]{mn2e}
\usepackage{graphicx,natbib}
\bibpunct{(}{)}{;}{a}{}{,}

\title{Map of the Galaxy in the 6.7 keV emission line}

\author[Revnivtsev, Molkov, Sazonov]{M.~Revnivtsev$^{1,2}$, S. Molkov
  $^{2,1}$, S. Sazonov$^{1,2}$\\
$^1$ Max-Planck-Institut f\"ur Astrophysik, Karl-Schwarzschild-Strasse 1, 85741
Garching, Germany\\
$^2$ Space Research Institute (IKI), Profsoyuznaya 84/32, Moscow 117997,
Russia\\
}

\pagerange{\pageref{firstpage}--\pageref{lastpage}}
\pubyear{2001}

\begin{document}
\maketitle

\label{firstpage}
\begin{abstract}
We study the two dimensional surface brightness distribution of the
Galactic X-ray background emission outside the central degree around
Sgr A$^*$ in the 6.7 keV line as measured by the PCA spectrometer of
the RXTE observatory. The use of the emission line instead of
continuum (3--20 keV) radiation and application of time variability
filtering to the long data set allows us to strongly suppress the
contamination of the GRXE map by bright point sources. The surface
brightness in the 6.7 keV line demonstrates very good correspondence
with the near-infrared surface brightness over the whole Galaxy,
supporting the notion that the GRXE consists mostly of integrated
emission from weak Galactic X-ray sources. We find compatible linear
correlations between near-infrared and 6.7 keV surface brightness
for the bulge and disk of the Galaxy. This indicates that the
populations of weak X-ray sources making up the GRXE in the disk and
bulge are not significantly different.

\end{abstract}

\begin{keywords}
radiation mechanisms: general--stars: binaries: general--Galaxy:
bulge--Galaxy: disc--Galaxy: general
\end{keywords}

%

\sloppypar

\section{Introduction}
Galactic ridge X-ray emission \cite[e.g.][]{cooke70,bleach72,
worrall82,warwick85} -- X-ray radiation concentrated to the Galactic
plane and unresolvable into bright ($>0.1-1$ mCrab)
point sources -- has a prominent spectral feature at energy $\sim6.7$ keV
\citep{koyama86,koyama89} that is typical of hot optically thin
plasma emission. As the energy resolution of X-ray detectors increased
more emission lines were found, additionally hinting at a thermal
origin of the GRXE.

The hypothesis of a truly diffuse origin of the GRXE has met a number of
 practically unresolvable difficulties \cite[see e.g.][for a
 review]{tanaka02}. The main problem is that the GRXE is apparently
 emission of optically thin plasma with temperature up to $>$5--10
 keV. Such hot diffuse plasma cannot be bound to the gravitational
 potential or magnetic field of the Galaxy and should form a continous
 outflow with a very large energy loss rate
($\sim 10^{43}$ erg/s). To sustain stationary X-ray extended emission, this
 energy must somehow be supplied throughout the whole Galaxy.

The alternative explanation of the GRXE being cumulative emission of
a large number of weak point X-ray sources emitting a strong 6.7 keV
line \cite[e.g.][]{worrall83,mukai93} has also faced
difficulties due to the failure of X-ray telescopes (including the
modern CHANDRA amd XMM-Newton) to resolve the
GRXE \citep{sugizaki01,hands04,ebisawa05}.

A solution to all these problems has apparently been found recently
through studies of the GRXE morphology. As the knowledge of the GRXE surface
brightness distribution in the Galaxy progressively improved \cite[e.g.][]
{koyama86,yamauchi93,yamauchi96,revnivtsev03,revnivtsev06}, it finally became
possible to demonstrate that the GRXE closely follows the near-infrared
emission of the Galaxy, which is in turn a good tracer of
the stellar mass density \citep{revnivtsev06}. It was consequently
proposed that X-ray emissivity is proportional to
stellar mass density. The determined unit-stellar-mass emissivity
proved to be in good agreement with the cumulative emissivity of X-ray
sources (cataclysmic variables and coronal stars) in the Solar neighborhood
\citep{sazonov06,revnivtsev06}. These findings imply that
the GRXE represents integrated emission of weak ($L_X<10^{34}$~erg/s)
Galactic X-ray sources.

Despite the significant progress made in understanding the GRXE
morphology, \cite{revnivtsev06} could not construct a two-dimensional
map of the GRXE of reasonable quality.
The main limiting factor was severe contamination by bright X-ray sources.
This in particular precluded a study of the GRXE distribution in the Galactic
plane. However, the prominent emission line at
energy $\sim6.7$ keV makes it possible to get rid of (or at
least strongly diminish) the contribution from bright sources. Indeed,
luminous X-ray binaries do not typically exhibit an emission line at
this energy. At most they show a fluorescent emission line at
$\sim6.4$  keV, but its equivalent width is approximately 10 times
smaller than that of the 6.7--6.9 keV emission lines
present in the GRXE. The contribution of the fluorescent line will
therefore be important only from very bright sources. The 6.7 keV line
has been successfully used in studying the GRXE by e.g. \cite{yamauchi93}
in application to GINGA/LAC data. Here we would like to apply the same
method to RXTE/PCA data, which provide better sensitivity and angular
resolution.

One can consider three distinct spatial components of the GRXE:
Galactic disk, Galactic bulge/bar, and the central cusp of extended
emission. The focus of the present paper will be on the former two
components, which are distributed widely over the sky spanning over
$100^\circ$ in Galactic longitude and $\sim3-8^\circ$ in Galactic
latitude. RXTE observations ideally suit for such a study.

The central surface brightness cusp of the GRXE
\citep[][]{koyama96,muno04,neronov05} has a size of only
5--10$^\prime$, meaning that X-ray instruments with good angular
resolution should be invoked to study this component. However, when
studying the GRXE on large scales one needs wide sky coverage and
large grasp (collecting solid angle multiplied by effective area) to be able to
detect flux from regions of low X-ray surface brightness.

\section{Analysis}

\begin{figure}
\includegraphics[width=\columnwidth,bb=31 188 570 415,clip]{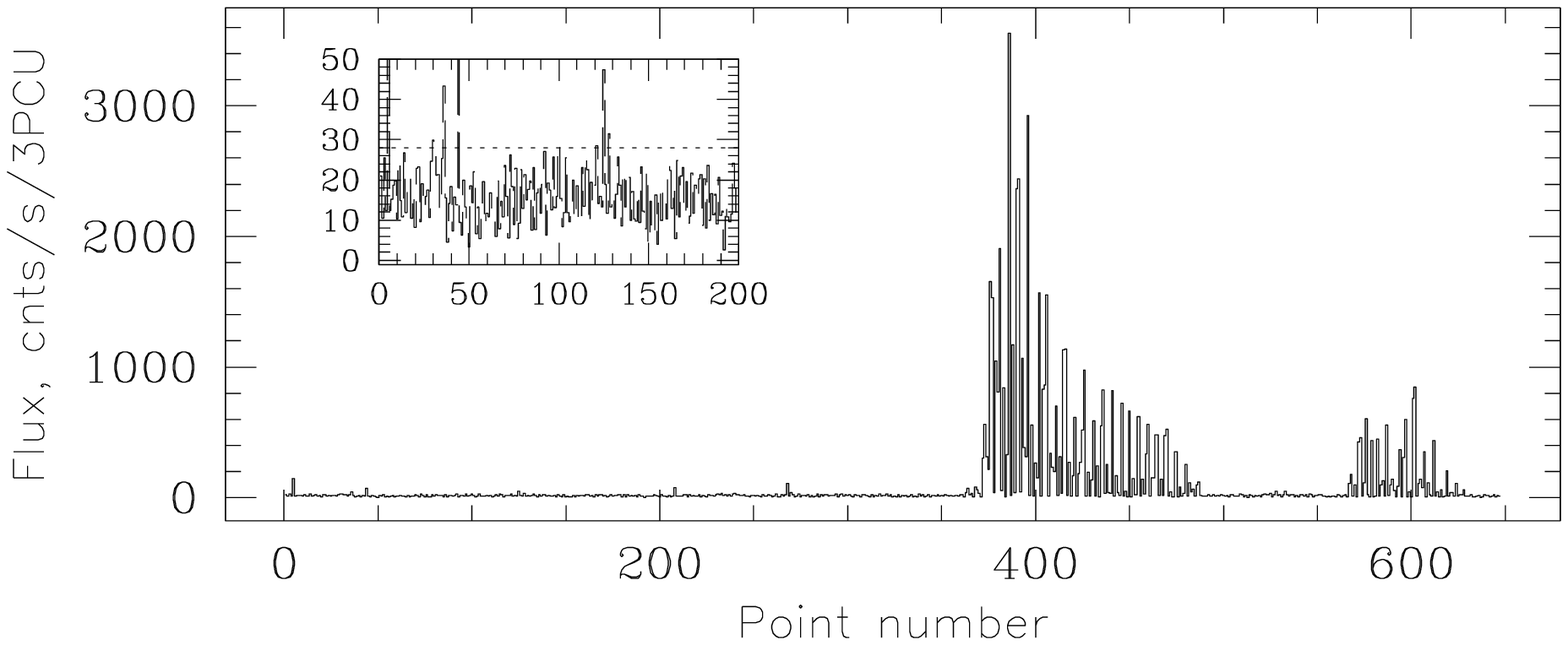}
\includegraphics[width=\columnwidth,bb=36 188 570 620,clip]{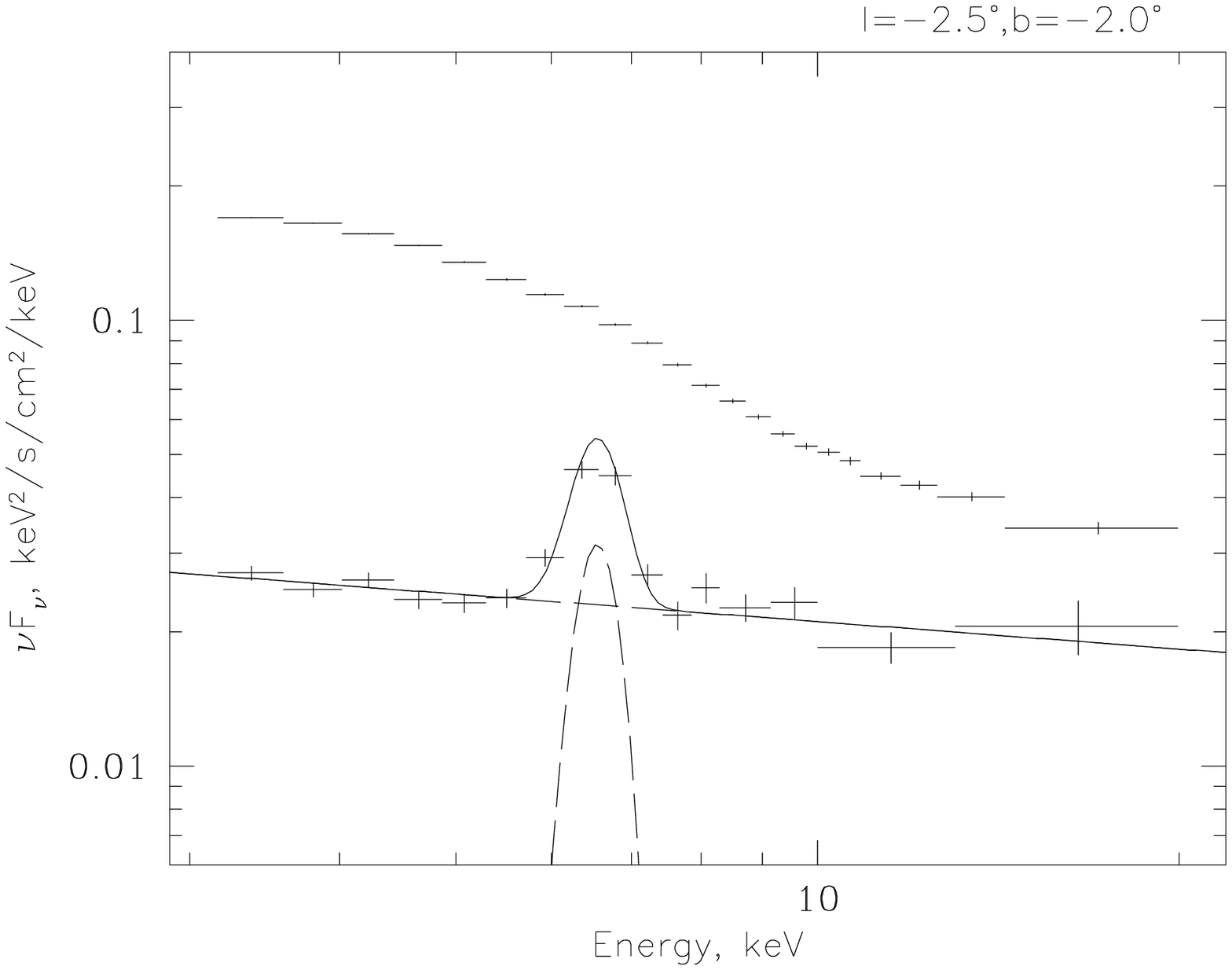}
\caption{{\bf Upper panel}: History of flux measurements made by
  RXTE/PCA within the $0.5^\circ\times0.5^\circ$ region centered at
  $l=-2.5,b=-2.0$. Strong
contamination by a transient bright X-ray source is obvious. In the inset
schematically shown is the flux thresholds that was used for data filtering.
{\bf Lower panel}: Energy spectrum measured by RXTE/PCA within the
 same region. The upper spectrum was obtained by collecting all data, while
the lower spectrum was obtained by collecting only data that passed
  the flux filtering as shown above. The lower spectrum is
  approximated by a model, which is later used
for extraction of emission line fluxes: a power law with a broad gaussian
emission line centered at $E_{\rm line}=6.66$ keV with a fixed width
  $\sigma=0.3$ keV}
\label{demo}
\end{figure}

\subsection{General approach}
In comparison with the work of \cite{revnivtsev06} the available RXTE data
allow us to make improvement in two directions.

\begin{itemize}
\item Flux and time filtering. Regular and numerous RXTE observations
of the Galactic bulge and inner Galactic disk regions \cite[e.g.][]{swank03}
have covered a long period of time, much longer than the typical
variability time scale of bright X-ray binaries. Therefore,
although the time averaged map of the inner Galaxy is dominated by the
emission of bright point sources (the large fraction of them being
transients), by filtering out time periods when bright sources are
present at particular positions on the sky it is possible to construct
a map that is almost free from contamination.

\item $\sim$6.7 keV emission line flux measurements. The GRXE contains
  lines of ionized iron at energies $\sim6.7-6.9$ keV
with a very large eqivalent width ($\sim$1 keV), which is not
typical of any type of luminous X-ray binaries.
The equivalent width of the emission line intrinsic
to the GRXE has been shown to be very stable across the Galactic disk
and bulge \cite[e.g.][]{tanaka02,revnivtsev03}.
For these reasons the 6.7 keV line may be regarded as a reliable
indicator of the GRXE.

\end{itemize}

\subsection{Data}
We use data of the PCA spectrometer of the
RXTE observatory. This instrument presents a number of advantages for
our study: it combines a large ($\sim6400$ cm$^2$) total effective
area with moderate energy resolution $\sim$18\% at energies 6--7 keV,
while its instrumental background is well understood and can be accurately
subtracted from the total detector count rate \citep[e.g.][]{craig02}.

We analyzed all observations performed in slew or scan mode from
March 1996 till March 2005. A large part of the data used is
associated with a series of dedicated Galactic bulge and plane scans
\footnote{More detailed information about the Galactic Center scans
  can be found at http://lheawww.gsfc.nasa.gov/
users/\,craigm/\,galscan/\,main.html}. These scans cover approximately a
square-like region $\sim15^\circ\times15^\circ$ around the Galactic Center
\citep[see e.g.][]{revnivtsev03} and two rectangular regions in
the inner Galactic plane. In total, these scans cover the inner
Galaxy from $l\sim-25^\circ$ to $l\sim+21^\circ$.  The
rest of the Galaxy is covered by occasional scans and slews.

\begin{figure}
\includegraphics[width=\columnwidth,bb=20 160 600 720,clip]{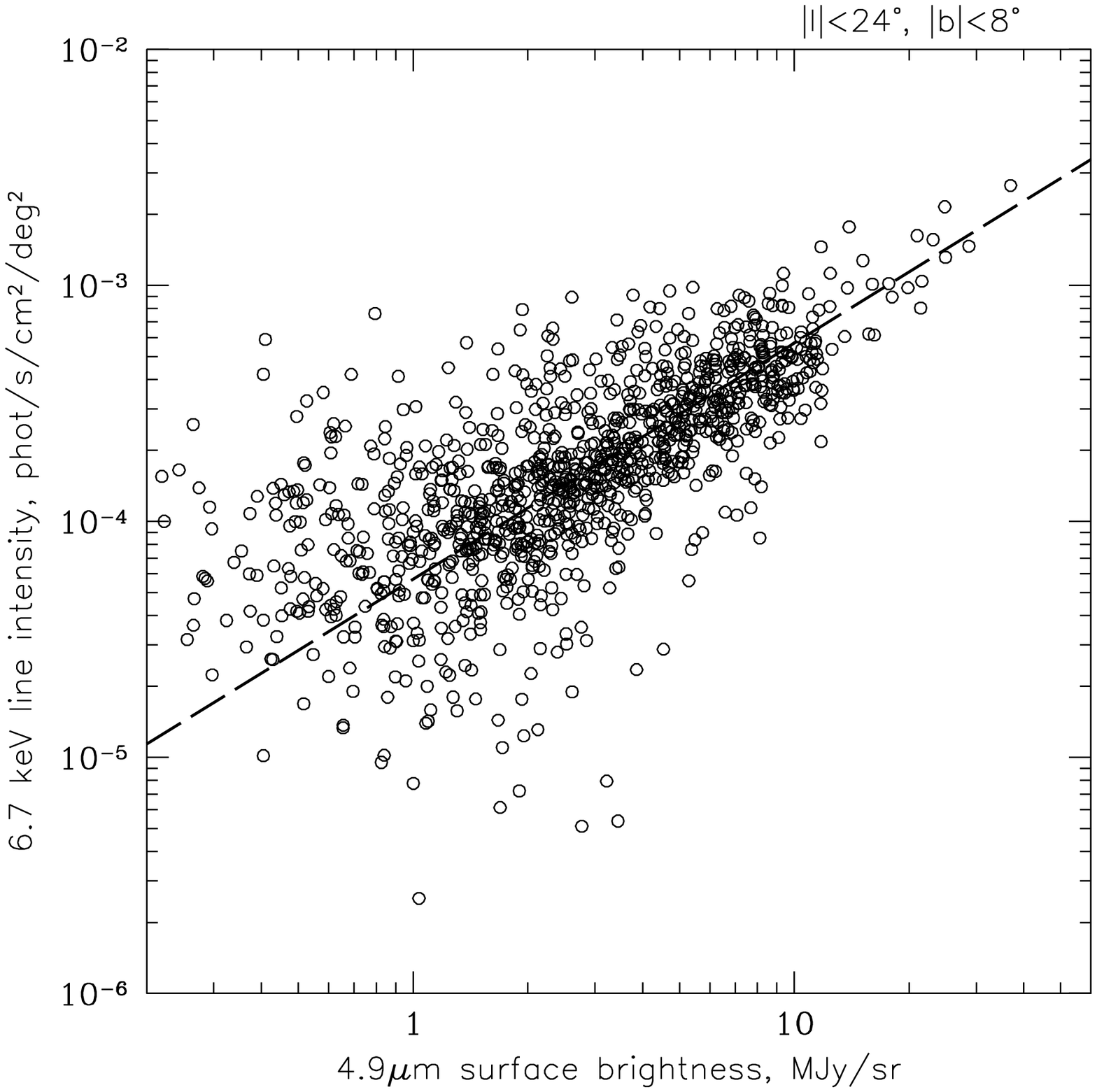}
\includegraphics[width=\columnwidth,bb=20 160 600 510,clip]{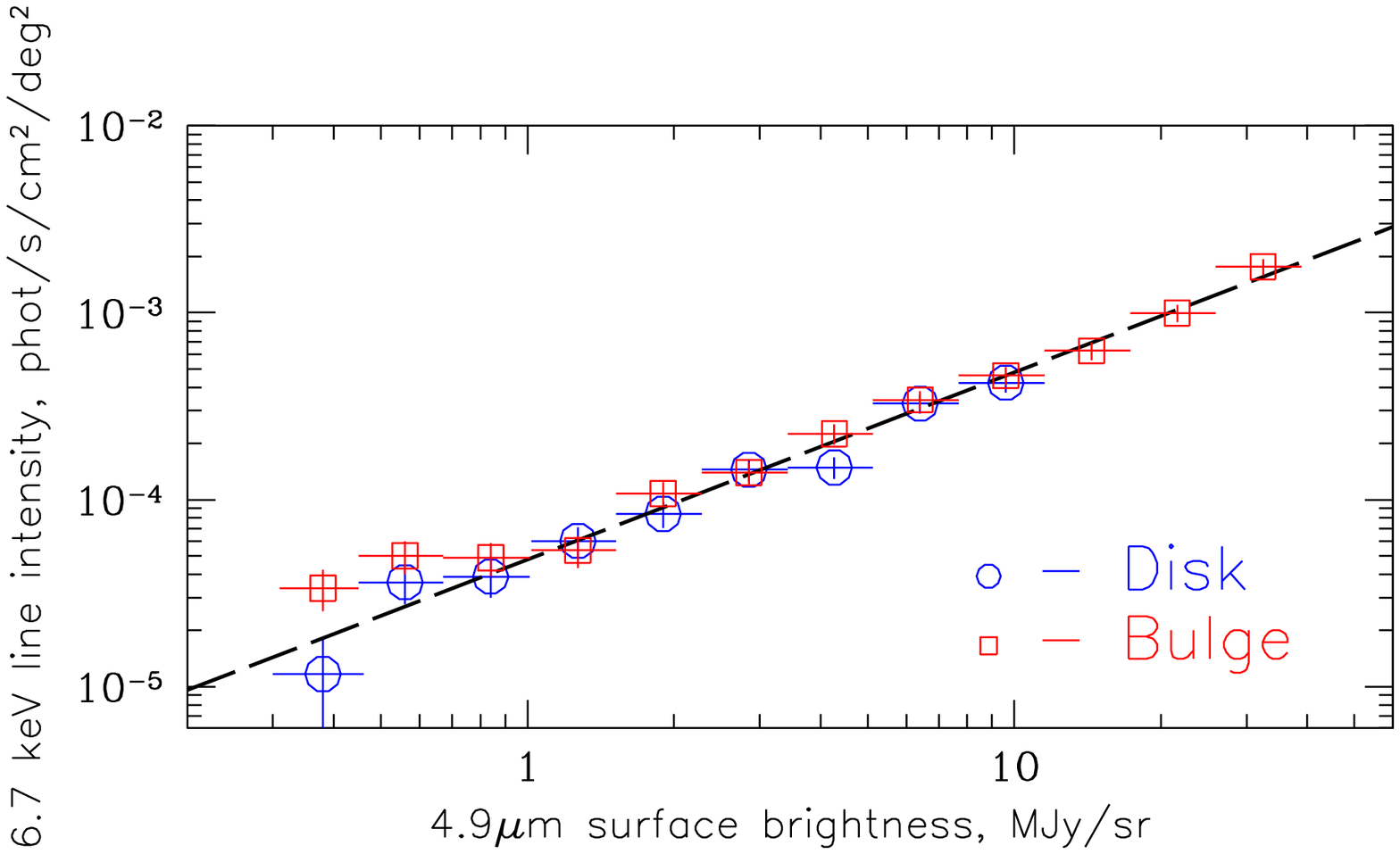}
\caption{Correlation of near-infrared (DIRBE $4.9\mu m$, corrected for
 interstellar reddening) and 6.7 keV emission line surface brightness
in the inner Galaxy. {\bf Upper panel:}
Points are measurements of X-ray and NIR intensities in small,
$0.5^\circ\times 0.5^\circ$, spatial bins, so that the X-ray
 statistics is rather limited. The uncertainties of the 6.7 keV line
fluxes are different for different points, but in general the scatter
 around the linear approximation is compatible with the errors
 involved. The correlation
$I_{\rm 6.7 keV line}$[phot/s/cm$^2$/deg$^2$]$=4.7\times10^{-5} I_{\rm
 4.9\mu m}$ [MJy/sr] is shown by the dashed line. {\bf Lower panel:}
 measurements of the X-ray line intensity averaged within NIR flux
 bins. Data for the Galactic disk ($|l|>10^\circ$) and bulge
 ($|l|<3^\circ$) are shown by open circles and squares,
 respectively. The dashed line is the same best-fit linear corelation
 as the one presented in the upper panel.
}
\label{scatter}
\end{figure}

The data were analyzed using standard tasks of the HEASOFT 6.0.2 package
(http://heasarc.gsfc.nasa.gov/docs/software/lheasoft/). Only Standard2 data
(129 energy channels, 16 s time resolution) from the first layers of
anodes of the detectors were used because of the higher quality
of the background subtraction possible for data from these layers. The total
exposure time of the observations used is $\sim30$Ms.

The data were rearranged into spatial bins of different size.
In the region $\sim20^\circ$ around the Galactic Center the bin size
was chosen to be  $0.5^\circ\times0.5^\circ$. Further away from the
Galactic Center the bin sizes were increased to $5^\circ
\times1^\circ$ (the longer dimension being along the Galactic plane).

Within these bins we searched for strong flux variability and selected
only those time intervals when the detected flux was around its most
probable value. Such data selection allows us to effectively filter
out observations that are significantly contaminated by transient
bright X-ray binaries. In order to demonstrate this we present in
Fig.~\ref{demo} the fluxes and spectra (the time averaged spectrum and the one
obtained after application of the time filter) measured by RXTE/PCA at
$l=-2.5^\circ, b=-2.5^\circ$.  It can be seen that in this particular
case time filtering has decreased the level of contamination by a bright
point source (X-ray Nova IGR J17464$-$3213) by an order of magnitude.

Upon applying time filtering we constructed the energy spectrum for every
spatial bin.

To approximate the spectral continuum we used the model of a
photoabsorbed power law
The centroid of the line was fixed at the value $E_{\rm line}=6.66$ keV
\citep[see e.g.][for the results of approximation of GRXE spectra measured
by instruments with moderate energy resolution]{koyama89,revnivtsev03}
and the width of the line was fixed at a value of $\sigma=0.3$
keV. Although such a representation of the complex of emission lines at
6.4--6.9 keV \cite[see e.g.][]{tanaka02} is clearly an
oversimplification, it provides a good approximation to the
observed set of lines for the limited energy resolution of
the PCA spectrometer (see Fig.\ref{demo}).

The flux in the broad gaussian emission line inferred from the spectral
approximation will be used below for mapping the GRXE.

Even after application of all our methods aimed at reducing the
contamination from bright
point sources, some spatial bins remain polluted. The strongest
contamination comes from  sources that themselves exhibit powerful
emission lines at 6.7--6.9 keV (e.g. thermal plasma emission).
In the Galactic Center region such sources are the Ophiuchus galaxy cluster
and the intermediate polar V2400 Oph (both are located at $l\sim0^\circ,
b\sim8^\circ$), in the Galactic plane strong contamination comes from
the star $\eta$ Carinae ($l\sim-75$). We excluded these spatial bins from
our subsequent analysis.

Below there will be presented measurements of the extended emission
in terms of surface brightness. For conversion of the flux measured by
PCA to intensity we assumed uniform surface brightness within the
PCA field of view ($\sim1^\circ$ radius) and adopted the solid angle of
the PCA collimator to be $0.975$ sq.deg. \citep{jahoda06}.

\subsection{Systematics}

The spectrometer PCA of the RXTE observatory is a well-calibrated instrument
\citep[see e.g.][]{jahoda06}, so no significant systematical problems
are expected during the data analysis. The accuracy of
the PCA instrumental background subtraction is very good,
$\sim1$\% of the average PCA background count rate,  or
$\sim1-2\times 10^{-12}$ erg/s/cm$^2$ in the energy band 3--20 keV. For
the GRXE this translates into a flux of the broad
($\sigma=0.3$ keV) gaussian line $dF_{\rm line}\sim5-8\times10^{-6}$
photons/s/cm$^2$/deg$^2$. Therefore, in our case this uncertainty
practically never dominates over poissonian count statistics.

There are two main uncertainties associated with our analysis: 1)
unfiltered contribution from bright point sources and 2) spatial
confusion due to the relatively fast ($\sim0.5-1$ deg per the 16-s
time bin) motion of the PCA field of view.

We managed to reduce the influence of the first problem by filtering out those
spatial bins for which the measured equivalent width of the 6.7 keV emission
line was significantly (more than twice) different from that measured
in the (high-quality) average GRXE spectrum by \cite{revnivtsev03}. Such
deviations of the equivalent line width indicate that the contribution of
non-GRXE emission is important.

The spatial confusion problem appears in those regions where the surface
brightness of the GRXE exhibits sharp features. In such cases, due to
the fast motion of the PCA field of view, the flux measured by PCA may
be ascribed to the wrong region. For PCA, with its
$\sim1^\circ$-radius field of view, this problem is particularly
important within a few degrees of Sgr A$^*$ because of the sharp
increase of the 6.7 keV line surface brightness toward it. Since our
analysis becomes unreliable within the $\sim1^\circ$ region around Sgr A$^*$,
we do not consider it here.

The confusion problem may also lead to distortions in our surface
brightness map near the Galactic plane due to the relatively large surface
brightness gradients in this region. Given that the exponential
vertical scale height of the GRXE around the Galactic plane is
$\sim1-2^\circ$ \citep[see e.g.][]{valinia98,revnivtsev06} and
assuming an angular velocity of the PCA field of view of $\sim1$
deg/16 s, we can estimate that the uncertainty of flux measurement
can reach $\sim10-15\%$. This uncertainty should be taken into account
in comparing the observed X-ray surface brightness with e.g. the
surface brightness of the Galaxy in the near infrared. In our subsequent
analysis we quadratically added a 10\% systematic uncertainty to allow
for confusion.

\subsection{NIR surface brightness data}

In the present paper we compare the GRXE map with the stellar
density distribution in the Galaxy for which the best tracer is
the near-infrared surface brightness.

The map of the Galaxy in the near-infrared spectral band was obtained
using data of COBE/DIRBE observations (the zodi-subtracted mission average
map provided by the LAMBDA archive of the Goddard Space Flight Center,
http://lambda.gsfc.nasa.gov). In order to reduce the influence of the
interstellar reddening we considered the DIRBE spectral band $4.9\mu$m.

We applied the simplest corrections to the NIR map of the Galaxy
obtained by COBE/DIRBE. We removed the extragalactic background component
determined by averaging measurements at $|b|>20^\circ$. We assumed
that the intrinsic NIR color temperature (i.e. the ratio of the intrinsic
surface brightnesses $I_{1.2\mu m}$ and $I_{4.9\mu m} $) of the
Galactic disk and bulge/bar is uniform and its true value can be
derived at high Galactic latitudes where interstellar reddening is
negligible. Then the foreground extinction map can be expressed as
\[
A_{\rm 4.9\mu m}={-2.5\over{A_{1.2\mu m}/A_{4.9\mu
m}-1}}\left[\ln\left({I_{1.2\mu m}\over{I_{4.9\mu
m}}}\right)-\ln\left({I_{1.2\mu m}^{0}\over{I_{4.9\mu
m}^{0}}}\right)\right].
\]

Here the $A$ values are the reddening coefficients at different
wavelengths. We used the interstellar reddening values from
\cite{lutz96,indebetouw05}. The applied correction of course
removed only the main effects of interstellar extinction on the COBE/DIRBE
map. Therefore, we do not expect that the obtained COBE/DIRBE map and
profiles have accuracy better than $\sim10\%$.

\section{Results}

\begin{figure}
\includegraphics[width=\columnwidth,bb=20 170 600 550,clip]{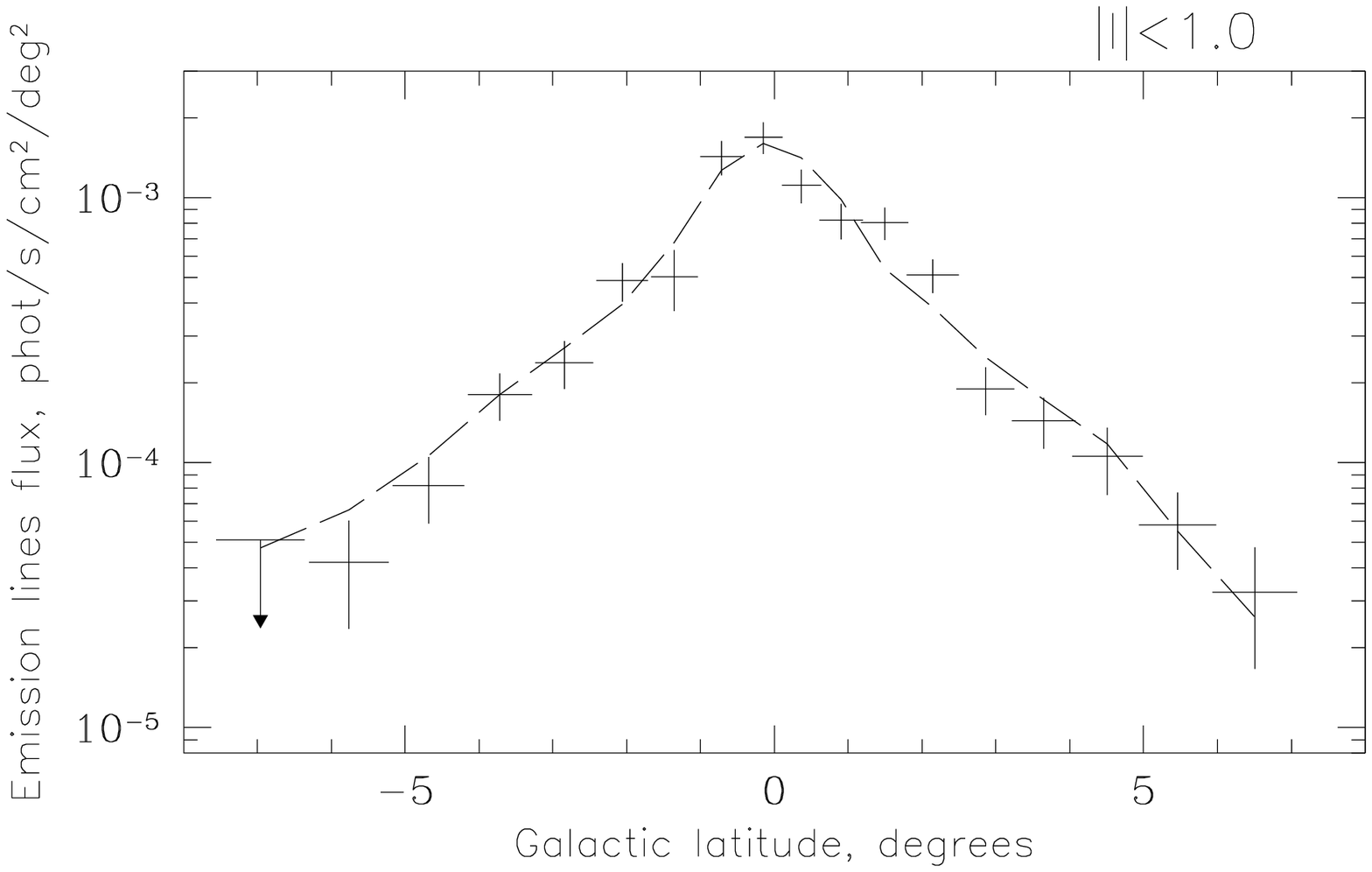}
\includegraphics[width=\columnwidth,bb=20 170 600 730,clip]{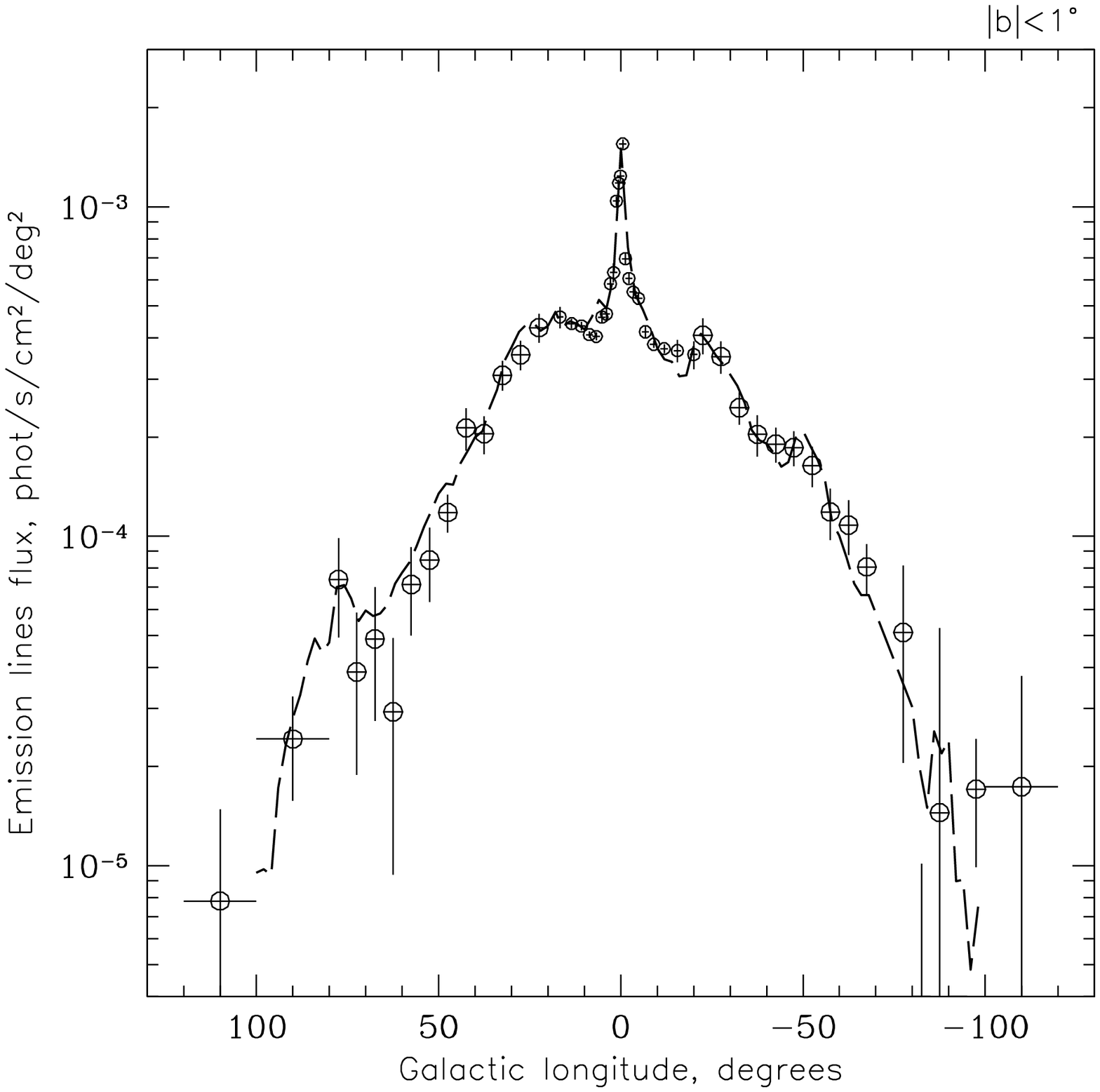}
\caption{Profiles of the surface brightness of the Galaxy in the 6.7
  keV emission line measured by RXTE/PCA. {\bf Upper panel:} Profile
  perpendicular to the Galactic plane at $|l|<2^\circ$. {\bf Lower
  panel:} Profile along the Galactic plane at $|b|<1^\circ$. On both
  panels the dashed lines show the profile of the surface brightness
  of the Galaxy at 4.9$\mu$m measured by COBE/DIRBE.}
\label{slice}
\end{figure}

In Fig. \ref{scatter} and \ref{slice} we demonstrate the correlation
between the 6.7 keV line intensity and the NIR surface brightness.
These results strongly support the finding of \cite{revnivtsev06} that
the GRXE surface brightness traces the NIR one.
With respect to that work, a significant improvement is achieved in
the Galactic plane region. It can be seen that the Galactic bulge and
plane regions
of the GRXE have the same proportionality to the NIR emission.
This indicates that there is no significant difference in the populations of
weak X-ray sources that constitute the GRXE in the plane and in the bulge.
Note that a similar result has been obtained via a study of the hard X-ray
emission of the Galactic ridge with INTEGRAL \citep{krivonos06}.

\begin{figure}
\includegraphics[height=\columnwidth,angle=-90]{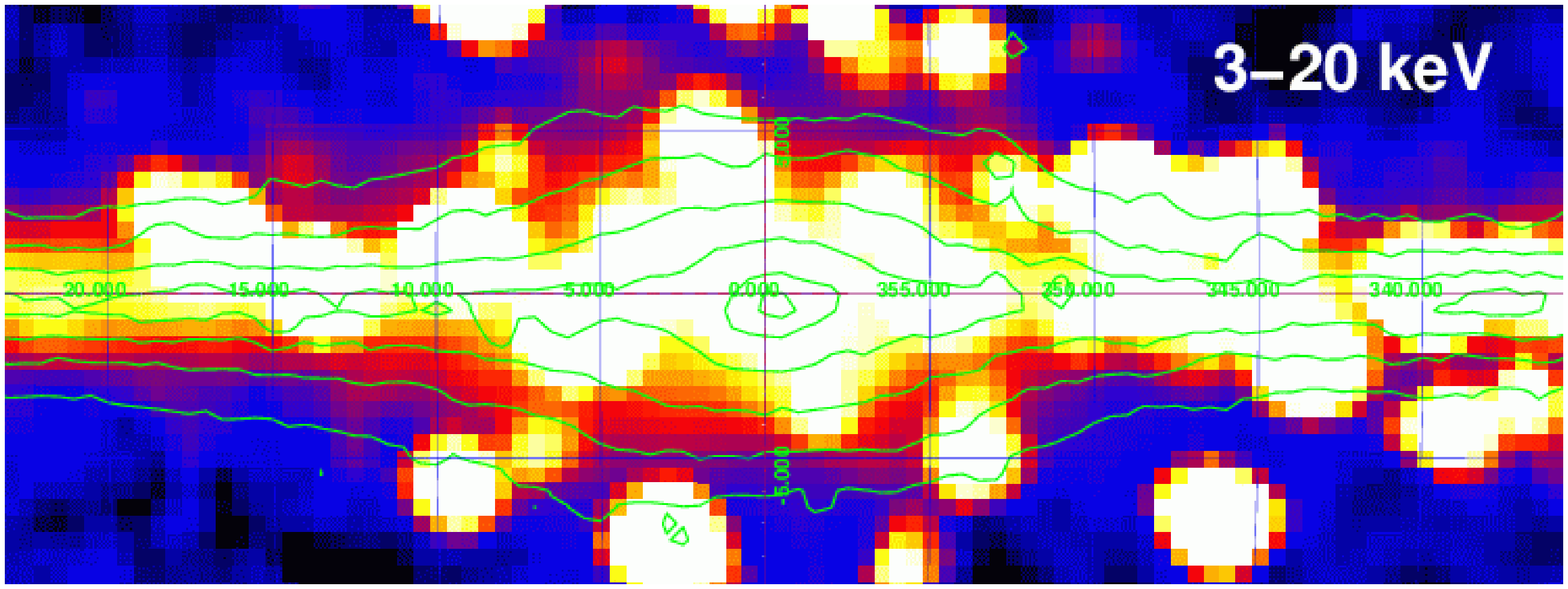}
\includegraphics[height=\columnwidth,angle=-90]{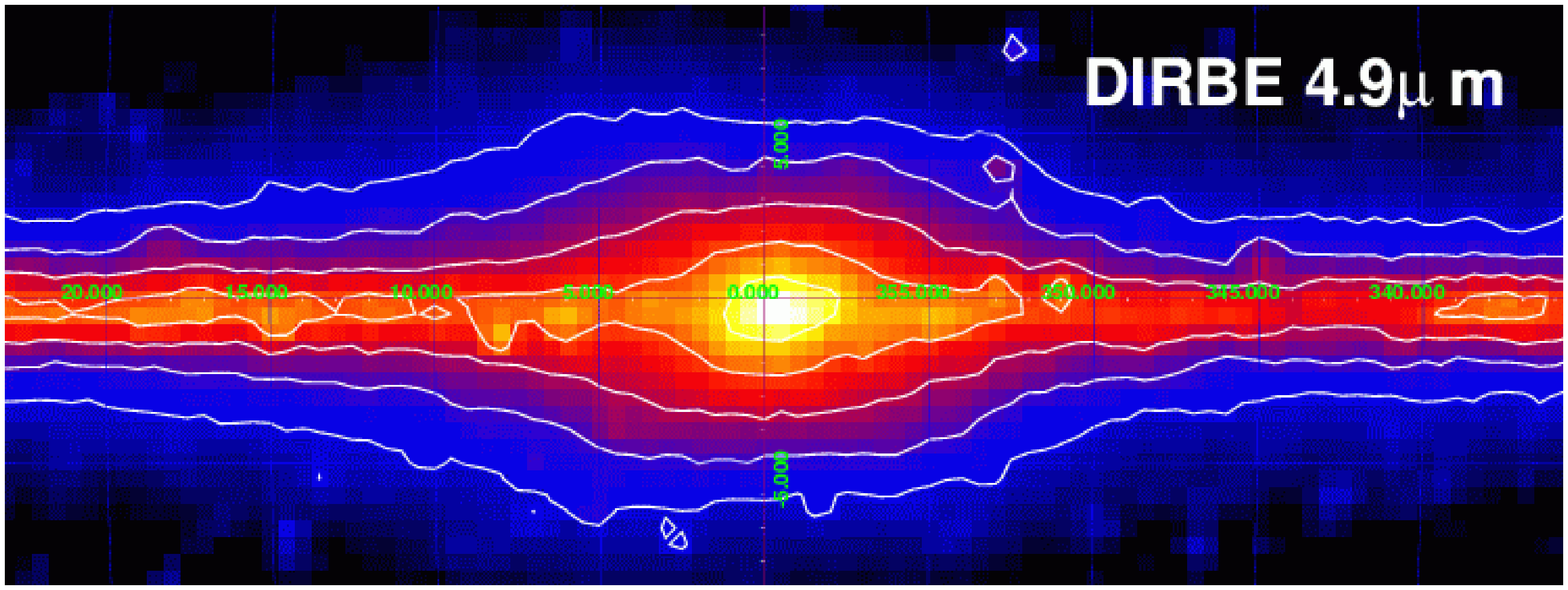}
\includegraphics[height=\columnwidth,angle=-90]{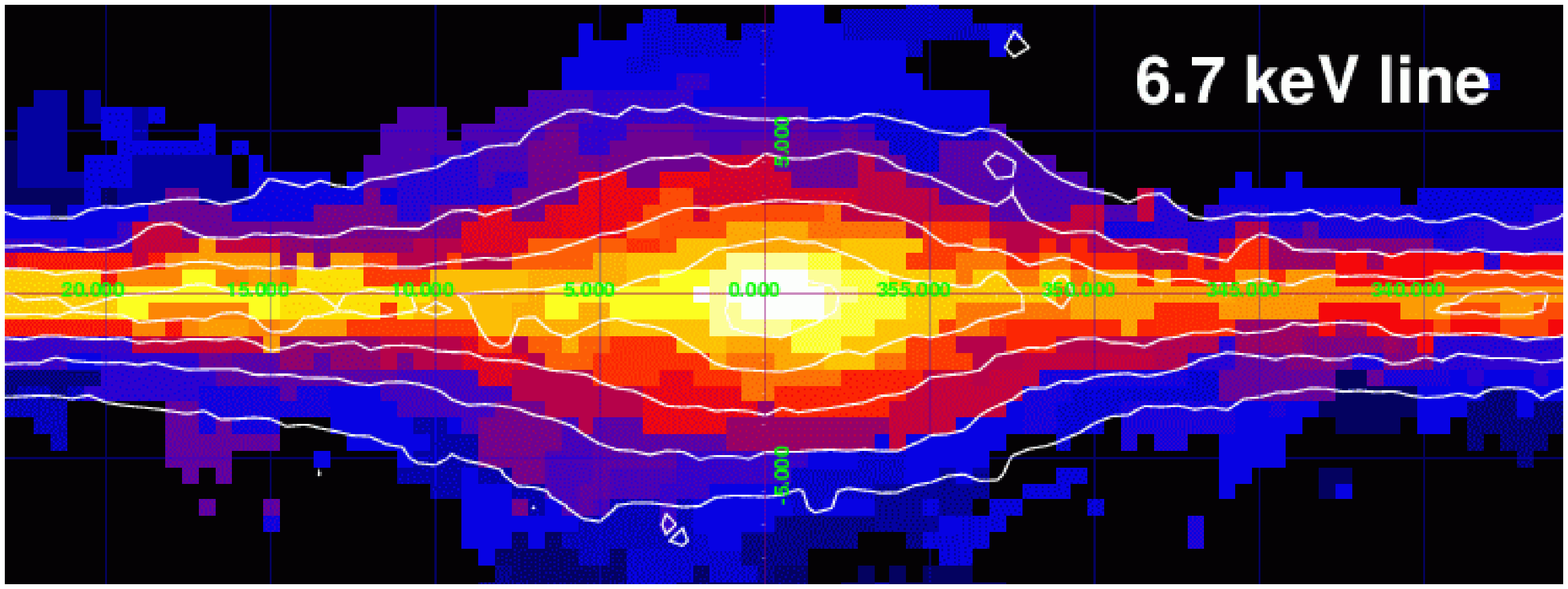}
\caption{{\bf Top:} Time averaged map of the inner Galaxy in the
  energy band 3--20 keV obtained with RXTE/PCA. The domination of bright
  point sources is evident. The contours are iso-brightness contours
  of the NIR emission of the Galaxy (see middle panel)
{\bf Middle:} Near-infrared surface brightness map of the Galaxy
  (COBE/DIRBE 4.9$\mu$m data, corrected for reddening)
{\bf Bottom:} Map of the surface brightness of the inner Galaxy in the
  6.7 keV emission line. The white contours are iso-brightness
  contours of NIR emission.}
\label{maps}
\end{figure}

In order to visualize the distribution of the 6.7 keV surface
brightness in the inner Galaxy we present a false-color
two-dimensional map in Fig. \ref{maps} and compare it with the maps
of Galactic emission in the 3--20 keV energy band and in the NIR band.
In constructing the 6.7 keV map we increased the statistics of flux
measurements in the 6.7 keV line by using adaptively sized spatial bins.

\section{Summary}

\begin{itemize}
\item We built a map of the Galaxy in the $\sim6.7$ keV line, the
characteristic emission line of the Galactic X-ray background (GRXE).
The use of only this line instead of a broad-band X-ray flux allowed
us to strongly suppress the contamination from bright point sources. As a
result we achieved a very good coverage of the inner Galaxy
($|l|<25^\circ$) and followed the Galactic ridge emission up to
$|l|\sim100-120^\circ$.

\item We demonstrated that the surface brightness of the Galaxy in the
6.7 keV line very well corresponds to its near-infrared surface brightness.
This shows that the GRXE volume emissivity is proportional
to the stellar mass density in the Galaxy.

\item We showed that the proportionality between the 6.7 keV surface
brightness and NIR one is the same for the bulge and
disk of the Galaxy.

\item The aforementioned observational facts along with the value of
unit-stellar-mass X-ray emissivity measured in the Solar neighborhood
\citep{sazonov06} provides further evidence that the bulk of the GRXE is made
up by faint Galactic X-ray point sources.

\end{itemize}

\label{lastpage}
\end{document}